\documentclass[11pt,a4paper]{article}
\usepackage[margin=1in]{geometry}
\usepackage{amsmath, amssymb, amsthm}
\usepackage{microtype}
\usepackage{hyperref}
\usepackage{enumitem}

\usepackage{empheq}
\usepackage{tikz}
\usepackage{subcaption}
\usepackage{comment}
\usepackage{tikz-cd}
\usetikzlibrary{arrows.meta,calc,positioning,decorations.markings,shapes.geometric}

\newcommand{\G}[2]{\Gamma(#1\leftarrow #2)}
\title{Phase information and measurement in the Stochastic--Quantum correspondence}
\author{Jason Doukas}
\date{}

\numberwithin{equation}{section}
\begin{document}
\maketitle

\begin{abstract}
The recently proposed stochastic–quantum correspondence establishes a connection between general stochastic processes and quantum dynamics, in which stochastic transition probabilities admit compatible representations on a Hilbert space, known as quantum lifts. We ask how the predictive information represented by off-diagonal degrees of freedom in the quantum lift is encoded within the stochastic description. We introduce the notion of a family of context-dependent stochastic processes and show that their continuation probabilities encode this predictive information. In particular, suitable continuations allow the intermediate lifted quantum state to be reconstructed entirely from system configuration probabilities. This perspective also clarifies the nature of quantum measurement: active continuations make basis-dependent quantum probabilities accessible as configuration probabilities, while a separate physical interaction records the realized configuration and introduces measurement disturbance. Although the rooted quantum probabilities alone do not guarantee that every stochastic realizer of the idealized recording interaction faithfully records the system's pre-measurement configuration, we show that at least one faithful stochastic realizer always exists.
\end{abstract}


\section{Introduction}
The stochastic--quantum correspondence establishes a connection between general stochastic processes and quantum dynamics \cite{Barandes2025,Barandes2025Indivisible,Doukas2026}, thereby providing the basis for a new ontic interpretation of quantum phenomena grounded in the conceptually transparent framework of general stochastic processes.

To describe this correspondence, let \(X=\{1,\ldots,N\}\) be a
finite configuration space and fix a root time \(t_0\in T\).
The stochastic dynamics are described by a classical dynamical
family rooted at \(t_0\), consisting of column-stochastic matrices
\begin{equation}
	\bigl\{\Gamma(t\leftarrow t_0)\bigr\}_{t\geq t_0},
\end{equation}
satisfying
\begin{equation}
	\Gamma(t_0\leftarrow t_0)=I.
	\label{eq:identityprop}
\end{equation}
A probability distribution at the root time is represented by a column vector \(p(t_0)\in\mathbb{R}^N\), with \(p_i(t_0)\geq0\) and \(\sum_i p_i(t_0)=1\), and evolves according to
\begin{equation}
	p(t)=\Gamma(t\leftarrow t_0)p(t_0).
\end{equation}

Such a classical dynamical family may arise from a family of stochastic processes parametrized by their initial probability distributions, collectively referred to as a stochastic realizer \cite{Doukas2026}.

The stochastic--quantum correspondence associates such a family with dynamics on an \(N\)-dimensional Hilbert space \(\mathcal H\). Let \(\{|i\rangle\}_{i=1}^N\) denote the distinguished configuration basis, and let
\begin{equation}
	\hat P^i = |i\rangle\langle i|
\end{equation}
be the corresponding configuration projectors. A compatible quantum lift is a family of completely positive trace-preserving (CPTP) maps \(\phi_{t,t_0}\) satisfying
\begin{equation}
	\Gamma(t\leftarrow t_0)_{ji}
	=
	\operatorname{Tr}\!\left[
	\hat P^{j}\,
	\phi_{t,t_0}(\hat P^{i})
	\right].
	\label{eqn:gammatrace}
\end{equation}
Thus, when the system is prepared in configuration \(i\) at the root time \(t_0\), the diagonal probabilities generated by the quantum lift reproduce exactly the stochastic transition probabilities \(\Gamma(t\leftarrow t_0)_{ji}\). As shown in \cite{Doukas2026}, every classical dynamical family admits a compatible quantum lift, and the class of such lifts encompasses all quantum channels as the underlying stochastic kernels vary.

The procedure for obtaining a stochastic description from a quantum system is formally straightforward: one fixes a basis, restricts to initially diagonal states, evolves them, and then projects onto the diagonal subalgebra. The resulting stochastic transition kernels form a rooted classical dynamical family describing the restriction of the quantum dynamics to a fixed configuration basis; see also \cite{Benatti2024}. This classical family admits a stochastic realizer.

The converse direction of the correspondence is also readily understood at a formal level. Since the evolution of probability distributions under a stochastic transition kernel is linear, it is natural that such dynamics admit representations on enlarged linear spaces---for example, by embedding probability vectors into the diagonal subalgebra of operators on a Hilbert space. 

The greater significance of the quantum lift, however, is that it may admit a division at the level of its dynamical maps even when the corresponding stochastic evolution is indivisible at the level of its two-time transition kernels. Such a division allows the future lifted state to be determined from the intermediate lifted state through a CPTP map. Consequently, a predictive-state description may exist at the level of the quantum lift even when the intermediate probability distribution alone cannot serve as a predictive state for future evolution, as is characteristically the case in non-Markovian processes.

A more striking feature, and one central to the present work, is that a stochastic description formulated relative to a fixed configuration basis can nevertheless support the phase-sensitive, no-preferred-basis measurement structure characteristic of quantum mechanics. At first sight, this is far from obvious. The compatibility condition defining a quantum lift \eqref{eqn:gammatrace} requires only that the projection onto the distinguished diagonal reproduce the configuration probabilities of the stochastic dynamics. One might therefore wonder whether the off-diagonal structure of the lifted quantum state is merely ornamental, with no independent operational significance in the underlying stochastic description.

As Barandes showed, this concern can be addressed operationally. By appropriately engineering the interaction between the system and a measuring device, the probabilities of the resulting device records can reproduce the familiar basis-dependent predictions of quantum mechanics for arbitrary choices of measurement basis. In this sense, the basis covariance of quantum theory emerges from an underlying stochastic process on a fixed configuration space.

If one takes seriously the proposal that all quantum phenomena arise from an underlying stochastic reality, a natural question arises: where, on the stochastic side, does quantum theory's rich phase-sensitive information reside?

A further question arises at the level of the stochastic description. Along a stochastic trajectory, the system possesses a definite configuration at each time. Merely conditioning on that realized configuration in one's description of the dynamics does not, in itself, constitute a physical disturbance of the system or alter its subsequent stochastic evolution. Quantum measurement, by contrast, generally alters subsequent predictions. The stochastic--quantum correspondence must therefore accommodate both definite underlying trajectories and the distinction between an unmeasured continuation and one involving a physical measurement.

The purpose of this paper is to clarify these two related issues. First, how is the phase-sensitive predictive information represented by the off-diagonal degrees of freedom of a quantum lift encoded in the stochastic description? Second, how does passive conditioning on a realized configuration differ from a physical measurement that records that configuration and thereby modifies the subsequent dynamics?

To address the first question, we introduce families of context-dependent stochastic processes and show that the predictive information carried by the off-diagonal degrees of freedom of a compatible, composable quantum lift is encoded in their continuation probabilities. Suitable continuations allow the intermediate lifted quantum state to be reconstructed entirely from configuration probabilities. To address the second question, we separate the active continuation used to expose basis-dependent quantum probabilities from the subsequent physical interaction that records a realized configuration. This makes it possible to distinguish, at the level of the stochastic process, the mathematical operation of conditioning on a realized configuration from the disturbance accompanying a physical recording interaction.

The paper is organized as follows. Section~2 develops the continuation-probability description of phase-sensitive predictive information, shows how suitable continuations allow the intermediate lifted state to be reconstructed from configuration probabilities, and introduces context-dependent stochastic realizers. It compares the parameter content of the quantum and stochastic descriptions and quantifies the remaining path-space freedom after the rooted quantum probabilities have been fixed. Section~3 treats measurement as a device recording interaction. It distinguishes active continuations from physical configuration recording, decomposes the idealized measurement interaction, establishes the existence of faithful stochastic realizers for the recording stage, and discusses observables and emergeables. A proof of the existence of a faithful realizer is given in the Appendix.

\section{Phase information}
\subsection{Phases from phase-blind dynamics}
\label{sec:phaseblind}
We begin by examining the question of quantum phase information within the stochastic description. 

\subsubsection{Stochastically indistinguishable unitaries}
Consider two unitaries $U_X$ and $U_Y$, with matrix elements taken in the configuration basis, such that they define the same stochastic matrix $\Gamma$\footnote{For example, on a two-dimensional Hilbert space, take
	\(U_X=e^{i\alpha\sigma_X}\) and \(U_Y=e^{i\alpha\sigma_Y}\),
	with the \(\sigma_Z\) eigenbasis as the configuration basis.
	Both induce
	\(\Gamma=
	\begin{psmallmatrix}
		\cos^2\alpha & \sin^2\alpha\\
		\sin^2\alpha & \cos^2\alpha
	\end{psmallmatrix}\).}:
\begin{equation}
	\Gamma_{ij} = p(i \mid j) = |(U_X)_{ij}|^2  =  |(U_Y)_{ij}|^2 .
\end{equation}
We call any unitaries sharing this property \emph{one-step stochastically indistinguishable}. Here ``indistinguishable'' means only that the associated one-step probabilities $p(i\mid j)$ on the single-system configuration space are identical.

Yet, despite being one-step stochastically indistinguishable, $U_X$ and $U_Y$ can yield different device-record probabilities once one allows the stochastic system to couple to a device. To see how probabilities usually associated with quantum measurements in a different basis can be obtained as configuration probabilities of a device, we use a stochastic formulation of the standard von Neumann system--device measurement interaction, following Barandes~\cite[Sec.~4.2]{Barandes2025}.

Consider a composite stochastic system consisting of a system $S$
and a device $D$. We assume that its classical dynamical family
is unistochastic and therefore admits a unitary lift. Throughout
this construction, we describe the stochastic dynamics using
their lifted Hilbert-space representation: unitary matrices
specify the corresponding unistochastic transition kernels,
while diagonal density matrices represent initial configuration
probability distributions.

Let the initial system distribution be $p_0$ and the device be
in its ready configuration $0$. In the associated Hilbert space,
with configuration bases $\{|s\rangle\}$ and $\{|d\rangle_D\}$,
the joint initial distribution is represented by
\begin{equation}
	\rho_{SD}(0)=\rho(0)\otimes|0\rangle_D\langle0|,
	\qquad
	\rho(0)=\sum_{s_0}p_0(s_0)|s_0\rangle\langle s_0|.
\end{equation}
For $U\in\{U_X,U_Y\}$, write $\rho=U\rho(0)U^\dagger$ for
the corresponding lifted system state.

Choose an orthonormal basis $\{|b_s\rangle\}$ with projectors $\hat P'{}^{s}:=|b_s\rangle\langle b_s|$. Let $W$ be the unitary satisfying $W|s\rangle=|b_s\rangle$, so that $\hat P'{}^{s}=W\hat P^sW^\dagger$. Choose device unitaries $R_s$ satisfying
\begin{equation}
	R_s|0\rangle_D=|d(s)\rangle_D,
\end{equation}
where the $d(s)$ are distinct device configurations. We specify the system--device measurement interaction by the joint unitary
\begin{equation}
	U_{SD}:=\sum_s\hat P'{}^{s}\otimes R_s.
	\label{eq:sys-dev-measure-int}
\end{equation}

The composite first-step configuration kernel is
\begin{equation}
	\Gamma_{SD}
	:=[U\otimes I_D]_{\odot}
	=\Gamma\otimes I_D.
\end{equation}
Here we use the shorthand $[A]_{\odot}:=A\odot A^{*}$ for the entrywise squared modulus of any matrix $A$, \(\odot\) denotes the Schur--Hadamard product \((X\odot Y)_{ij}:=X_{ij}Y_{ij}\), in the configuration basis, and \(^{*}\) denotes entrywise complex conjugation. 

The device is unchanged during this step, and summing over its final configurations recovers the common system transition kernel $\Gamma$.

The composite configuration kernel after the system--device interaction is
\begin{equation}
	\Gamma_{SD}'
	:=\bigl[U_{SD}(U\otimes I_D)\bigr]_{\odot}.
	\label{eq:USD-composite-int}
\end{equation}
These kernels specify a unistochastic classical dynamical family on the composite configuration space. As established in \cite{Doukas2026}, a stochastic realizer for this family always exists. In the product configuration basis,
\begin{equation}
	\begin{aligned}
		[\Gamma_{SD}']_{(j,d(s)),(s_0,0)}
		&=
		\left|
		\langle j,d(s)|
		U_{SD}(U\otimes I_D)
		|s_0,0\rangle
		\right|^2\\
		&=
		\left|\langle j|\hat P'{}^{s}U|s_0\rangle\right|^2.
	\end{aligned}
\end{equation}
The device probabilities are therefore
\begin{equation}
	\begin{aligned}
		p_D(d(s))
		&=
		\sum_{s_0}p_0(s_0)\sum_j
		[\Gamma_{SD}']_{(j,d(s)),(s_0,0)}\\
		&=
		\operatorname{Tr}\left(\hat P'{}^{s}U\rho(0)U^\dagger\right)
		=
		\operatorname{Tr}\left(\hat P'{}^{s}\rho\right).
	\end{aligned}
\end{equation}

Thus the configuration probabilities of the device in this composite stochastic process reproduce the projective quantum measurement probabilities associated, in the lifted description, with a chosen ``basis'' that need not coincide with the configuration basis. Whenever $U_X\rho(0)U_X^\dagger$ and $U_Y\rho(0)U_Y^\dagger$ differ for an allowed diagonal preparation, a suitable choice of measurement basis, implemented through the system--device interaction $U_{SD}$ entering Eq.~\eqref{eq:USD-composite-int}, gives different device probabilities. A single system transition kernel $\Gamma$ is therefore compatible with distinct stochastic processes that yield \emph{different} device-record probabilities, reproducing the phase-sensitive measurement predictions of the lifted quantum description.

If attention is restricted to this single kernel, however, the quantum description appears to carry more predictive information than the kernel itself. From this restricted perspective, one might be tempted to regard quantum mechanics as more fundamental, with $\Gamma$ merely a lossy “shadow” obtained by restricting preparation and readout to the configuration basis. If this conclusion is resisted, in the present two-unitary example the issue becomes concrete: what additional stochastic structure, beyond the one-step kernel $\Gamma$, carries the information represented by these distinct quantum evolutions?

\subsubsection{System continuations and predictive information}

In the construction above, these distinctions become visible through device configuration probabilities in a composite stochastic process. One might then think that the additional predictive information resides only in this larger process containing the apparatus. If so, accounting for these basis-dependent quantum predictions, usually attributed to the system alone, would require a stochastic description that included the combined system and device. The properties attributed to the system would then depend on the external apparatus and its interaction with the system. This conclusion, however, does not necessarily follow.

To see why, note that $\Gamma$ captures only \emph{one-step} information and that subsequent continuations of the system dynamics can reveal predictive distinctions that this kernel does not capture. Consider what happens at the next step, when a further unitary $V$ acts on the system alone. If the first-step realization is $U_X$, we obtain the two-step kernel
\[
\Gamma_1=[VU_X]_{\odot},
\]
whereas if the first-step realization is $U_Y$, we obtain
\[
\Gamma_2=[VU_Y]_{\odot}.
\]

Although $[U_X]_{\odot}=[U_Y]_{\odot}=\Gamma$, these two-step kernels need not coincide: a common continuation $V$ can reveal distinctions that are invisible in the one-step transition probabilities.\footnote{The surviving ambiguity consists of right multiplication by a unitary diagonal in the configuration basis. If \(U_Y=U_XD\) for such a \(D\), then the two unitaries act identically on every diagonal input and cannot be distinguished by any subsequent common continuation \(V\).}
If the Hadamard bracket respected composition (i.e., if $[AB]_{\odot}=[A]_{\odot}[B]_{\odot}$ for all unitaries $A$ and $B$), this would be impossible, since both kernels would equal $[V]_{\odot}\Gamma$. In general, however, this identity fails. The response to $V$ can therefore depend on features of the first-step realization that the common stochastic kernel $\Gamma$ does not record.

In the lifted description, the evolution is governed by composable CPTP maps on $\mathcal B(\mathcal H)$ that are unitary in the present example. For a fixed continuation $V$, the next step takes only the \emph{present} operator $\rho(t_1)$ as input; earlier configurations do not enter separately. To reproduce the different two-step predictions, the lifted description must therefore carry the distinguishing information forward in $\rho(t_1)$ through composition. For the same diagonal initial state, the two intermediate operators $U_X\rho(0)U_X^\dagger$ and $U_Y\rho(0)U_Y^\dagger$ have identical diagonals. Consequently, whatever distinguishes their responses to the common continuation must reside in their phase-sensitive, off-diagonal structure. This is where the lift carries the predictive information that is invisible in the common one-step stochastic kernel $\Gamma$.

Where does the corresponding distinction enter the stochastic description? By the existence result in \cite{Doukas2026}, we may choose a stochastic realizer for each of the two quantum evolutions considered above.

Using the chain rule, Eq.~(2.3) of Ref.~\cite{ Doukas2026}, we can write the two-step transition probabilities for the process associated with $U_X$ as 
\begin{align}
	[\Gamma_1]_{x_2,x_0}
	&=
	\sum_{x_1}
	p_1(x_2\mid x_1,x_0) p(x_1\mid x_0)
	\nonumber\\
	&=
	\operatorname{Tr}\left(
	\hat P^{x_2}V
	\left(U_X\rho_{x_0}U_X^\dagger\right)V^\dagger
	\right),
	\label{eq:twotimeconstrint}
\end{align}
where $\rho_{x_0}=\hat P^{x_0}$. An analogous expression holds for the process associated with $U_Y$ with $\Gamma_2$, $p_2$, and $U_Y$ in place of $\Gamma_1$, $p_1$, and $U_X$.

In the lifted description, the conjugation by $V$ is the same in both cases; only the intermediate state $U_X\rho(0)U_X^\dagger$ or $U_Y\rho(0)U_Y^\dagger$ can differ. On the stochastic side, the common kernel $\Gamma$ fixes $p(x_1\mid x_0)$, but does not determine the conditional continuation probabilities $p_1(x_2\mid x_1,x_0)$ and $p_2(x_2\mid x_1,x_0)$. Whenever the two-step kernels differ, these probabilities must therefore differ as well. Thus, the conditional continuation probabilities encode the different responses of the intermediate lifted states to the same continuation $V$.

These conditional probabilities may also depend on the earlier configuration $x_0$ beyond the present configuration $x_1$. Such history dependence is the familiar notion of \emph{memory} in a non-Markovian process.\footnote{As shown in \cite[Section~4(b)]{ Doukas2026}, this memory is unavoidable for some quantum systems: a revival of configuration-space distinguishability rules out every Markovian realizer on the same configuration space.}

\subsection{State reconstruction from configuration probabilities}

The configuration probabilities generated by continuations can do more than distinguish two candidate evolutions. By varying the continuation while keeping the preparation and preceding evolution fixed, one can recover all information about the intermediate lifted state from the resulting system configuration probabilities. We show that suitable continuations can generate system configuration probabilities equal to those that orthodox quantum theory would assign to positive-operator-valued measure (POVM) measurement outcomes. In our construction, these are probabilities of configurations realized by the stochastic trajectory; no measuring device or physical recording interaction is introduced. This provides a form of state tomography in the stochastic process setting, based on the system configuration probabilities.

In what follows, we use the mathematical tools of quantum measurement theory, interpreting the quantum description as a lift of the underlying stochastic dynamics.

\subsubsection{Configuration probabilities in POVM form}
Concretely, we specify an \emph{active} system continuation through a CPTP map on the lifted state:
\begin{equation}
	\Lambda(\rho)=\sum_{\beta}\Lambda_{\beta}\rho
	\Lambda_{\beta}^{\dagger},
	\qquad
	\sum_{\beta}\Lambda_{\beta}^{\dagger}\Lambda_{\beta}=I.
\end{equation}

This continuation induces the configuration probabilities
\begin{align}
	p(j)
	&=\operatorname{Tr}\bigl(\hat P^{j}\Lambda(\rho)\bigr)
	=\sum_{\beta}\operatorname{Tr}
	\bigl(\hat P^{j}
	\Lambda_{\beta}\rho\Lambda_{\beta}^{\dagger}\bigr),
	\label{eq-povm-continuation-probabilities}
\end{align}
which, by cyclicity of the trace, can be written as
\begin{equation}
	p(j)=\operatorname{Tr}(E_j\rho),
\end{equation}
where
\begin{equation}
	E_j:=\sum_{\beta}\Lambda_{\beta}^{\dagger}
	\hat P^{j}\Lambda_{\beta}.
\end{equation}
These operators satisfy $E_j\geq0$ and $\sum_jE_j=I$, so $\{E_j\}$ forms a POVM.

Here the $p(j)$ are the system configuration probabilities after the continuation, expressed in terms of the operators $E_j$ and the state $\rho$ before the continuation. These probabilities have the same mathematical form as POVM measurement probabilities in orthodox quantum theory. However, the projectors $\hat P^j$ appearing in the definition of $E_j$ originate from $\operatorname{Tr}(\hat P^j\Lambda(\rho))$, where they serve solely as mathematical tools for extracting the configuration probabilities of the stochastic process at that time. Evaluating this expression does not implement a physical measurement; it leaves the lifted state after the continuation, $\Lambda(\rho)$, unchanged.

As a special case, we can generate system configuration probabilities equal to those that orthodox quantum theory assigns to a projective measurement in the chosen orthonormal basis $\{|b_s\rangle\}$. Taking the single Kraus operator $\Lambda_1=W^\dagger$ for the basis rotation introduced above gives the active continuation $\Lambda(\rho)=W^\dagger\rho W$. The induced effects are then
\begin{equation}
	E_s=W\hat P^sW^\dagger=|b_s\rangle\langle b_s|=\hat P'{}^{s}.
\end{equation}
Substituting these effects into the probability formula above gives $p(s)=\langle b_s|\rho|b_s\rangle$. Here $p(s)$ is the probability that the stochastic trajectory occupies configuration $s$ after the active continuation, and is equal to the Born probability for outcome $s$ in the chosen measurement basis. Varying the unitary continuation therefore reproduces the \emph{basis-dependent} probabilities of projective quantum measurements as system configuration probabilities.

\subsubsection{Informationally complete continuations}
We now seek continuations whose configuration probabilities, taken together, contain enough information to reconstruct the intermediate lifted state.

We can always choose a CPTP continuation whose configuration probabilities equal those assigned by orthodox quantum theory to any prescribed POVM with at most \(N\) outcomes. For a POVM $\{E_j\}_{j=1}^{M}$ with $M\leq N$, one such map is
\begin{equation}
	\Lambda(\rho)=\sum_{j=1}^{M}
	\operatorname{Tr}(E_j\rho)\hat P^j.
	\label{eq:POVM-readout-channel}
\end{equation}
This map is CPTP, and the configuration probabilities after applying this continuation are $\operatorname{Tr}(\hat P^j\Lambda(\rho))=\operatorname{Tr}(E_j\rho)$ for $j=1,\ldots,M$.

Choose a POVM $\{E_j^{(c)}\}_{j=1}^{M_c}$ for each continuation context $c=1,\ldots,r$, with $M_c\leq N$, whose effects are collectively \emph{informationally complete}, meaning
\begin{equation}
	\operatorname{span}_{\mathbb R}
	\left\{
	E_j^{(c)}:
	c\in\{1,\ldots,r\},\;
	j\in\{1,\ldots,M_c\}
	\right\}
	=\operatorname{Herm}(\mathcal H),
	\label{eq:readout-informational-completeness}
\end{equation}
where $\operatorname{Herm}(\mathcal H)$ denotes the real vector space of Hermitian operators on $\mathcal H$.

Equation~\eqref{eq:POVM-readout-channel} then provides a CPTP continuation $\Lambda^{(c)}$ for each context $c$, whose configuration probabilities equal the outcome probabilities of the corresponding POVM. These probabilities are
\begin{equation}
	p_c(j)
	=\operatorname{Tr}\bigl(\hat P^j\Lambda^{(c)}(\rho)\bigr)
	=\operatorname{Tr}(E_j^{(c)}\rho).
\end{equation}
Because the effects span $\operatorname{Herm}(\mathcal H)$, these probabilities uniquely determine $\rho$.\footnote{Since
	$\operatorname{span}_{\mathbb R}\{E_j^{(c)}: c, j\}
	=\operatorname{Herm}(\mathcal H)$, we can select $N^2$ of the original effects
	\(
	F_\alpha=E_{j_\alpha}^{(c_\alpha)}, \alpha=1,\ldots,N^2,
	\)
	that form a basis of $\operatorname{Herm}(\mathcal H)$.
	Thus $\rho=\sum_{\beta=1}^{N^2}r_\beta F_\beta$ with real
	coefficients, and the corresponding outcome probabilities obey
	\[
	p_\alpha=\operatorname{Tr}(\rho F_\alpha)
	=\sum_{\beta=1}^{N^2}G_{\alpha\beta}r_\beta,
	\qquad
	G_{\alpha\beta}:=\operatorname{Tr}(F_\alpha F_\beta).
	\]
	The Gram matrix $G$ is invertible because the $F_\alpha$
	are linearly independent. Hence $r=G^{-1}p$, giving
	\[
	\rho=\sum_{\alpha,\beta=1}^{N^2}
	F_\alpha(G^{-1})_{\alpha\beta}p_\beta.
	\]} Keeping the preparation and preceding evolution fixed while varying only the continuation therefore yields system configuration probabilities that, taken together, determine the intermediate lifted state.

\subsection{Context-dependent stochastic continuations}

\subsubsection{Context dependent transition kernels}

In the preceding discussion, we have made use of the freedom to vary the continuation of a stochastic process, referring loosely to the choice of continuation as its context. We now refine this notion and suggest a physical interpretation of its origin.

In any experiment, we employ a variety of apparatus, probes, controls, and other devices that interact with the system under investigation and influence its dynamics. We abstract away these distinctions and regard the experimental setup simply as a configuration of the system's environment. At a microscopic level, such a configuration can be described in terms of the types, positions, and velocities of the particles constituting the environment. We refer to the different possible configurations of the environment at a given instant as \emph{environmental settings}.

We refer to an environmental setting operating over a fixed time interval as an \emph{experimental procedure}. A sequence of such procedures is referred to as an \emph{experimental context}.

We now express the configuration probabilities produced by the active continuations described above in terms of these context dependent stochastic transition kernels.

Consider three times $t_0<t_1<t_2$. Let $c_1$ denote the environmental setting specifying the experimental procedure operating between $t_0$ and $t_1$, and $c_2$ the setting specifying the procedure  operating between $t_1$ and $t_2$. The experimental context is therefore $(c_1,c_2)$.

The first procedure is represented in the lifted description
by a CPTP map $\Lambda^{(c_1)}$, giving the intermediate state
\begin{equation}
	\rho_{c_1}(t_1)
	=
	\sum_\alpha
	\Lambda^{(c_1)}_\alpha
	\rho(t_0)
	\Lambda^{(c_1)\dagger}_\alpha.
\end{equation}
Its induced phase-blind kernel is
\begin{equation}
	\Gamma_{c_1}(t_1\leftarrow t_0)
	=
	\sum_\alpha[\Lambda^{(c_1)}_\alpha]_{\odot}.
\end{equation}

Applying the second procedure $\Lambda^{(c_2)}$ to the
intermediate lifted state \emph{without first discarding
	its off-diagonal terms} gives the cumulative transition
kernel
\begin{equation}
	\Gamma_{c_1,c_2}(t_2\leftarrow t_0)
	=
	\sum_{\beta,\alpha}
	[\Lambda^{(c_2)}_\beta
	\Lambda^{(c_1)}_\alpha]_{\odot}.
\end{equation}

For a diagonal initial preparation $\rho(t_0)=J(p_0)$,
the cumulative stochastic kernel maps $p_0$ to the system
configuration probability vector
$\Gamma_{c_1,c_2}(t_2\leftarrow t_0)p_0$.
Its components are the probabilities in
Eq.~\eqref{eq-povm-continuation-probabilities},
now expressed in terms of stochastic transition data.

In particular, for an initial configuration $x_0$,
\begin{equation}
	p_{c_1,c_2}(x_2\mid x_0)
	=
	\operatorname{Tr}\left(
	E_{x_2}^{(c_2)}
	\Lambda^{(c_1)}(\hat P^{x_0})
	\right),
\end{equation}
where
\begin{equation}
	E_{x_2}^{(c_2)}
	=
	\sum_\beta
	\Lambda^{(c_2)\dagger}_\beta
	\hat P^{x_2}
	\Lambda^{(c_2)}_\beta.
\end{equation}
Thus, even for a fixed second procedure $c_2$, the
final configuration probabilities can depend on the
preceding procedure $c_1$ through the intermediate
lifted state.

On the stochastic side, each experimental context
$(c_1,c_2)$ is represented by a stochastic process
$p_{c_1,c_2}$. The conditional continuation
probabilities satisfy
\begin{equation}
	p_{c_1,c_2}(x_2\mid x_0)
	=
	\sum_{x_1}
	p_{c_1,c_2}(x_2\mid x_1,x_0)
	p_{c_1}(x_1\mid x_0).
\end{equation}
Here the first-step probabilities depend only on the
first procedure $c_1$, whereas the conditional
continuation probabilities may depend on both
procedures, as well as on the preceding configuration
history.

This distinction is illustrated by the two-unitary
example considered above. Let $c_X$ and $c_Y$ denote
the experimental procedures corresponding to $U_X$
and $U_Y$, respectively. Since these unitaries are
one-step stochastically indistinguishable,
\begin{equation}
	p_{c_X}(x_1\mid x_0)
	=
	p_{c_Y}(x_1\mid x_0)
	=:p(x_1\mid x_0).
\end{equation}

Now choose the same second procedure $c_V$,
corresponding to the unitary continuation $V$.
The resulting cumulative kernels are
\begin{equation}
	\begin{aligned}
		\Gamma_{c_X,c_V}(t_2\leftarrow t_0)
		&=[VU_X]_{\odot},\\
		\Gamma_{c_Y,c_V}(t_2\leftarrow t_0)
		&=[VU_Y]_{\odot}.
	\end{aligned}
\end{equation}
As shown above, these kernels need not coincide,
even though the first-step stochastic kernels
are identical. Consequently, whenever their final
probabilities differ, the conditional continuation
probabilities
$p_{c_X,c_V}(x_2\mid x_1,x_0)$ and
$p_{c_Y,c_V}(x_2\mid x_1,x_0)$ must differ for
at least some configuration histories.

More generally, the first procedure fixes
$p_{c_1}(x_1\mid x_0)$, while the second specifies
how that evolution is continued. Within each context,
the conditional probability
$p_{c_1,c_2}(x_2\mid x_1,x_0)$ may depend on the
earlier configuration $x_0$ beyond the present
configuration $x_1$, as well as on both experimental
procedures.

Varying the second procedure $c_2$ while keeping
the first $c_1$ fixed provides different ways
of exposing the predictive information retained
from the preceding evolution: when the corresponding
effects are informationally complete, the resulting
probabilities determine the intermediate density
operator $\rho_{c_1}(t_1)$. The information represented
by that operator is therefore accessible through the
configuration probabilities of a family of
context-dependent stochastic processes, which the
first-step kernel
$\Gamma_{c_1}(t_1\leftarrow t_0)$ alone does not
determine.

\subsubsection{Sequences of continuations}

We now extend the construction above from two
successive experimental procedures to a sequence
of such procedures, with an environmental setting
chosen at each step.

Consider sampling times $t_0<t_1<\cdots<t_m$.
Suppose the same $r$ experimental procedures corresponding to CPTP maps 
$\{\Lambda^{(c)}\}_{c=1}^{r}$ are available at every step.
At step $k$, the setting $c_k\in\{1,\ldots,r\}$
selects the procedure $\Lambda^{(c_k)}$.
The full experimental context is the sequence
$\mathbf c=(c_1,\ldots,c_m)$.

Writing $\mathbf c[1:k]:=(c_1,\ldots,c_k)$, the
cumulative map up to step $k$ is
\begin{equation}
	\Phi_{\mathbf c[1:k]}
	:=
	\Lambda^{(c_k)}\circ\cdots\circ\Lambda^{(c_1)}.
\end{equation}

For each experimental context, let
\begin{equation}
	\Gamma_{\mathbf c[1:k]}(t_k\leftarrow t_0)
	:=
	\sum_{\beta_1,\ldots,\beta_k}
	\left[
	\Lambda^{(c_k)}_{\beta_k}\cdots
	\Lambda^{(c_1)}_{\beta_1}
	\right]_{\odot}
\end{equation}
denote the rooted kernel induced by the cumulative map.

\paragraph{Context-dependent stochastic realizers.}
A \emph{context-dependent stochastic realizer} of
these kernels is specified by context-dependent
joint laws of the form
\begin{equation}
	p_{\mathbf c,p_0}(x_0,\ldots,x_m)
	=
	p_0(x_0)\prod_{k=1}^{m}
	p(x_k\mid x_{k-1},\ldots,x_0;c_1,\ldots,c_k)
\end{equation}
that reproduce the prescribed rooted probabilities:
\begin{equation}
	p_{\mathbf c,p_0}(x_0=i,x_k=j)
	=
	p_0(i)
	[\Gamma_{\mathbf c[1:k]}(t_k\leftarrow t_0)]_{ji}
\end{equation}
for every $k$, experimental context $\mathbf c$,
and initial distribution $p_0$. The conditional laws
$p(x_k\mid x_{k-1},\ldots,x_0;
\allowbreak c_1,\ldots,c_k)$ remain fixed as $p_0$
varies and depend on the experimental procedures
only through the prefix $(c_1,\ldots,c_k)$,
so they do not depend on future settings.

Such a realizer always exists. The endpoint
construction \cite{Doukas2026} gives one by choosing
\begin{equation}
	p(x_k\mid x_{k-1},\ldots,x_0;c_1,\ldots,c_k)
	=
	[\Gamma_{\mathbf c[1:k]}(t_k\leftarrow t_0)]_{x_kx_0}.
\end{equation}
Each fixed experimental context and initial
distribution therefore determines a stochastic
process from this common conditional structure.

\subsection{Parameter counts and path-space freedom}

We now examine the amount of information that can be carried by context-dependent stochastic realizers, measured by their real degrees of freedom. We first compare the numbers of parameters needed to specify context-dependent composable quantum evolutions and general context-dependent stochastic path laws. We then quantify the freedom remaining in a stochastic realizer after the rooted quantum predictions have been fixed.

\subsubsection{Parameter counts for quantum and stochastic evolutions}
In discrete time with $m$ steps on $N$ configurations and a fixed, full-support initial distribution $p_0$, an arbitrary unitary description involves at most $m\dim U(N)=mN^2$ real degrees of freedom\footnote{Similarly, with one CPTP map per time step, the real degrees of freedom remain polynomial in $N$ and $m$, 	namely $m(N^4-N^2)$.}, whereas a completely general stochastic process is specified by a joint law on $X^{m+1}$ with this initial marginal and hence requires $N^{m+1}-N$ independent real parameters.\footnote{Neither count includes the initial distribution $p_0$. Including it adds $N-1$ independent real parameters to both descriptions.}

To include different experimental contexts, $\mathbf c=(c_1,\ldots,c_m)$, fix the same initial distribution $p_0$ in all contexts. At step $k$, setting $c_k$ selects a unitary $U^{(c_k)}$ from the $r$ available unitaries, independently of the preceding settings. The cumulative unitary up to step $k$ is $U^{(c_k)}\cdots U^{(c_1)}$.

We count the real parameters in the step-by-step specification of these cumulative evolutions, counting the matrix parameters separately for each setting at each step. This specification therefore involves at most
\begin{equation}
	mrN^2
\end{equation}
real parameters.

On the stochastic side at step $k$, there are $r^k$ setting sequences $(c_1,\ldots,c_k)$, with $N^k$ possible configuration histories $(x_0,\ldots,x_{k-1})$ for each setting sequence. Each pair specifies a conditional distribution, $p(x_k\mid x_{k-1},\ldots,x_0;c_1,\ldots,c_k)$,  over the $N$ possible configurations $x_k$, requiring $N-1$ independent real parameters after normalization. The family of joint laws requires these conditional distributions at every step, so summing over $k$ gives
\begin{equation}
	(N-1)\sum_{k=1}^{m}(Nr)^k
\end{equation}
independent real parameters.

Thus, the parameter count for the general stochastic family grows exponentially with the number of steps, whereas the count for the composable quantum description grows only linearly.

From this viewpoint, the original question---``how can a phase-blind stochastic description encode the phase-sensitive information of quantum mechanics?''---has the emphasis backwards. A general context-dependent stochastic path law can carry extensive information through conditional probabilities that depend on both preceding configurations and chosen settings. Rather, the question is how a composable quantum lift, when one exists, can retain the relevant predictive distinctions while being specified by comparatively few real parameters.

The key is that the lift does not encode the full stochastic path law. It reproduces the specified transition probabilities and the configuration probabilities under the allowed contexts. Many inequivalent stochastic realizers can reproduce these same probabilities while differing in their multi-time joint distributions.

In this restricted sense, the phase structure of the lifted state can be viewed as a compressed memory of the preparation and preceding context history. Its off-diagonal components can preserve predictive distinctions that are absent from the instantaneous configuration distribution but affect configuration probabilities under subsequent continuations. In a composable evolution these distinctions are carried forward in the present state, allowing subsequent predictions to be computed without separately specifying the preparation and preceding context-dependent evolution.

\subsubsection{Freedom after fixing the quantum predictions}
The rooted transition probabilities reproduced by the lift need not uniquely determine the underlying trajectory law. To quantify this freedom, fix a continuation context $c$ and its prescribed rooted kernels $p(x_1\mid x_0)$ and $p_c(x_2\mid x_0)$. We count the degrees of freedom in the three-time laws compatible with these two kernels. Eq.~\eqref{eq:twotimeconstrint} requires
\begin{equation}
	p_c(x_2\mid x_0)
	=
	\sum_{x_1}
	p_c(x_2\mid x_1,x_0) p(x_1\mid x_0).
	\label{eq:fixed-rooted-kernel-constraint}
\end{equation}
The quantities being varied are the higher-order conditionals $p_c(x_2\mid x_1,x_0)$, while the context and both rooted kernels remain fixed.

For each fixed $x_0$ (there are $N$ choices), the conditional $p_c(x_2\mid x_1,x_0)$ may be regarded as an $N\times N$ stochastic matrix in the indices $(x_2,x_1)$: for each $x_1$ it specifies a probability distribution over $x_2$. Before imposing the constraint in Eq.~\eqref{eq:fixed-rooted-kernel-constraint}, it therefore has $N(N-1)$ degrees of freedom for each $x_0$, and across all $x_0$ this amounts to
\[
N\cdot N(N-1)=N^3-N^2
\]
free parameters.

Fixing the two rooted kernels imposes the constraint in Eq.~\eqref{eq:fixed-rooted-kernel-constraint} for every pair $(x_2,x_0)$. For each $x_0$ these give $N$ equations in $x_2$, of which $N-1$ are independent due to normalization. Thus there are $N(N-1)=N^2-N$ independent constraints in total. Consequently, when both prescribed kernels have strictly positive entries, the space of admissible three-time conditionals has
\[
(N^3-N^2)-(N^2-N)=N(N-1)^2
\]
remaining degrees of freedom.

Thus, even within a single fixed context, the prescribed rooted transition probabilities can leave substantial freedom in the underlying multi-time stochastic law.

\section{Measurement as a device recording interaction}
\subsection{Configuration recording}
It is important to distinguish the role of the projectors appearing in the above POVM constructions from their role in a physical measurement. The expression
\begin{equation}
	p(j)=\operatorname{Tr}\bigl(\hat P^j\Lambda(\rho)\bigr)
	\label{eq:configuration-readout-probability}
\end{equation}
extracts the probability that the stochastic trajectory occupies configuration $j$ at that time. It identifies these probabilities without introducing a measuring interaction or changing the lifted state $\Lambda(\rho)$; in particular, it does not imply that a physical dephasing operation has occurred.

From the stochastic perspective, the trajectory already has a realized configuration, whether or not an observer knows it. Learning which configuration is realized requires a physical interaction that encodes it in a device record.

We introduced an idealized measurement interaction in Eq.~\eqref{eq:sys-dev-measure-int} at the beginning of this section. Choosing the configuration projectors $\hat P^s:=|s\rangle\langle s|$ in that interaction gives the \emph{configuration-recording unitary}:
\begin{equation}
	C_{SD}:=\sum_s\hat P^s\otimes R_s.
	\label{eq:configuration-recording}
\end{equation}

For a sharp configuration input, the unitary satisfies
\begin{equation}
	C_{SD}\bigl(|s\rangle\otimes|0\rangle_D\bigr)
	=
	|s\rangle\otimes|d(s)\rangle_D.
\end{equation}

The stochastic shadow of the configuration-recording interaction, $M_{SD}:=[C_{SD}]_{\odot}$, has entries
\begin{equation}
	[M_{SD}]_{(s,d),(r,d_0)}
	=
	\delta_{s,r}
	\left|\langle d|R_r|d_0\rangle\right|^2,
\end{equation}
which, for an initially ready device, reduces to
\begin{equation}
	[M_{SD}]_{(s,d),(r,0)}
	=
	\delta_{s,r}\delta_{d,d(r)}.
	\label{eq:recording-appendix-kernel}
\end{equation}

For sharp configuration preparations with the device ready, the kernel specifies the deterministic map $(r,0)\mapsto(r,d(r))$: the system configuration is preserved while the device acquires its record.

\subsection{Decomposing the measurement interaction}
The configuration-recording interaction is a special case of the more general idealized measurement interaction. Using it, we can describe the general measurement interaction in Eq.~\eqref{eq:sys-dev-measure-int} in successive stages. 

Let the measurement occur at $t_m$, and let $t_m^-$ and $t_m^+$ denote the times immediately before and after the measurement interaction. 

The incoming state is 

\begin{equation}
	\rho^{(1)} = \rho(t_m^-).
\end{equation}

The active continuation stage gives the state immediately before configuration recording
\begin{equation}
	\rho^{(2)}:=\Lambda(\rho^{(1)}),
\end{equation}
resulting in configuration probabilities at this stage given by $\operatorname{Tr}(\hat P^s\Lambda(\rho^{(1)}))$.

When $C_{SD}$ is applied after the active continuation $\Lambda$, it produces the device record $d(s)$ with probability $\operatorname{Tr}(\hat P^s\Lambda(\rho^{(1)}))$. Therefore the device records inherit the same configuration probabilities as the system. The outcome-averaged reduced state after configuration recording becomes
\begin{equation}
	\rho^{(3)}
	:=
	\operatorname{Tr}_D \left[
	C_{SD}\bigl(\rho^{(2)}\otimes|0\rangle_D\langle0|\bigr)
	C_{SD}^{\dagger}
	\right]
	=
	\sum_s\hat P^s\rho^{(2)}\hat P^s.
\end{equation}

Thus, although the configuration recording interaction preserves the instantaneous system configuration probabilities, it removes the off-diagonal components of the reduced state and can therefore change subsequent configuration probabilities.

From the stochastic perspective, this shows that the disturbance arises from the physical interaction used to acquire information about the trajectory, rather than being required for the configuration itself to be realized.

For a unitary continuation $\Lambda(\rho)=W^\dagger\rho W$, the first stage applies the active rotation $W^\dagger$ to the system. The subsequent configuration-recording interaction $C_{SD}$ records $d(s)$ with probability $\operatorname{Tr}(\hat P'{}^{s}\rho^{(1)})$. This reproduces the outcome probabilities of a projective quantum measurement in the primed basis, while leaving the system in the sharp \emph{configuration} state $|s\rangle$ conditional on this record. To place the system in the corresponding eigenstate $|b_s\rangle$ of the measured observable, as in an ideal projective quantum measurement, one must apply an additional active rotation $W$ after the configuration recording interaction, without changing the device.

In this unitary case, the final rotation acts through the inverse channel $\Lambda^\dagger$, giving
\begin{equation}
	\rho^{(4)}:=W \rho^{(3)} W^\dagger
	=\rho(t_m^+).
\end{equation}

The procedure we have just described performs the physical measurement in three stages. The original unitary measurement interaction \eqref{eq:sys-dev-measure-int} combines these stages into a single joint interaction, producing the same outcome probabilities and final conditional system states $|b_s\rangle$. This relation is expressed by the decomposition

\begin{equation}
	U_{SD}
	=
	(W\otimes I_D) C_{SD} (W^\dagger\otimes I_D).
	\label{eqn:USD_CSD_decomp}
\end{equation}

This decomposition makes explicit two distinct roles:
\begin{enumerate}
	\item \emph{Active basis rotation.}
	The system-only operation $\rho^{(1)}\mapsto\rho^{(2)}=W^\dagger\rho^{(1)} W$  makes the desired measurement-basis probabilities available 	as configuration probabilities:
		\begin{equation}
		p(s)
		=
		\operatorname{Tr}(\hat P^s\rho^{(2)})
		=
		\operatorname{Tr}(\hat P'{}^{s}\rho{(t_m^-)}).
	\end{equation}
	This rotation creates no device record. After recording, the final rotation $W$ maps the conditional configuration state $|s\rangle$ to $|b_s\rangle$.
	
	\item \emph{Disturbance due to configuration recording.} The interaction $C_{SD}$ produces the device record $d(s)$ and removes the off-diagonal components of the outcome-averaged reduced state in the configuration basis. This is the disturbance associated with the recording step.
\end{enumerate}

This separation clarifies the role of the measuring device in recovering phase-sensitive predictions. The basis-dependent measurement outcome probabilities need not be attributed solely to an enlarged stochastic process containing the apparatus. Differences in the measurement outcome probabilities arise from the chosen active continuation on the system, while the same configuration-recording interaction can be used across measurement settings.

\subsection{Faithful recording in stochastic realizers}
\label{sec:faithful-recording}
The rooted classical dynamical family corresponding to the system--device evolution with the measurement interaction admits a stochastic realizer, as shown in \cite{ Doukas2026}.

In what follows, we take $\Lambda=\mathrm{id}$ without loss of generality, absorbing the active continuation into the preceding evolution. The recording interaction then consists only of $C_{SD}$. Write $S_-,D_-$ and $S_+,D_+$ for the system and device configurations at $t_m^-$ and $t_m^+$, respectively. For the input $\rho(t_m^-)$ and a ready device, the measurement interaction prescribes
\begin{equation}
	\begin{aligned}
		p_+(s,d)
		&:=\langle s,d|C_{SD}
		\bigl(\rho(t_m^-)\otimes|0\rangle_D\langle0|\bigr)
		C_{SD}^{\dagger}|s,d\rangle\\
		&=\langle s|\rho(t_m^-)|s\rangle\delta_{d,d(s)}.
	\end{aligned}
	\label{eq:recording-appendix-output-probabilities}
\end{equation}
Equation~\eqref{eq:recording-appendix-output-probabilities} ensures that the device agrees with the final configuration, $D_+=d(S_+)$, with probability one. It does not establish that $S_+=S_-$, or that the device records the system configuration realized immediately before the interaction. An example of a realizer that flips the incoming system configuration and records the final one while reproducing the prescribed rooted probabilities is given in Appendix~\ref{app:intermediate-recording}.

We call a realizer \emph{faithful} across the recording interval if it preserves the incoming configuration and records it in the device. With the device initially ready, this means that, on conditioning events of positive probability,
\begin{equation}
	\Pr(S_+=s,D_+=d\mid S_-=r,D_-=0)
	=
	\delta_{s,r}\delta_{d,d(r)}.
	\label{eq:faithful-recording}
\end{equation}
We show in Appendix~\ref{app:intermediate-recording} that such a realizer can always be chosen.

\subsection{Observables and emergeables}

Thus far, we have been concerned with configurations and their probabilities. We now briefly describe the physical values associated with these configurations and their device records. A configuration observable is a random variable represented by $A:=\sum_s a_s\hat P^s$. Its values and physical units are part of the physical specification of the system. The assignment $d(s)\mapsto a_s$ associates each device record with the corresponding observable value. The rotated projectors $\hat P'{}^{s}$ introduced above, together with the same values $a_s$, define the observable $\widetilde A_W:=WAW^\dagger=\sum_s a_s\hat P'{}^{s}$. When $\widetilde A_W$ is non-diagonal in the configuration basis, it is called an \emph{emergeable}~\cite{Barandes2025}. For a fixed readout assignment, the same device configuration $d(s)$ may therefore represent outcomes of different emergeables for different choices of $W$.

The configuration observables are distinguished from the emergeables in that their values are determined by the underlying ontic configuration of the model. Yet, as we have seen, this ontic construction gives rise to a Hilbert-space description whose predictions are preserved under consistent changes of basis. In practice, this same basis covariance obscures which physical observables correspond to the underlying ontic configuration variables~\cite{Barandes2025Indivisible}. Thus the construction supports an ontic description at the level of the model without identifying those variables with the fundamental configuration variables of nature.

\section{Conclusion}

The central result of this work is that the phase-sensitive predictive information represented by the off-diagonal degrees of freedom of a quantum lift has a counterpart in the underlying stochastic description. Although a single stochastic transition kernel does not contain this information, it is encoded in the conditional continuation probabilities of a family of context-dependent stochastic processes. By varying the experimental procedures, these predictive distinctions can be exposed through configuration probabilities alone. Indeed, suitable informationally complete procedures allow the intermediate lifted quantum state to be reconstructed entirely from such probabilities. In this sense, the phase information of the quantum description is not merely ornamental, but reflects predictive information present in the stochastic description.

This perspective also clarifies the distinction between conditioning on a realized configuration and physically measuring it. In a stochastic process, the configuration is definite irrespective of whether it is observed, and conditioning on its value does not itself constitute a physical disturbance. Physical measurement, by contrast, requires an interaction with a recording device. We have shown how this interaction can be decomposed into an active continuation, which makes arbitrary-basis quantum probabilities accessible as configuration probabilities, and a recording interaction, which produces a device record and introduces disturbance. Although the rooted quantum probabilities do not guarantee that every stochastic realizer faithfully records the pre-measurement configuration, at least one faithful realizer always exists for the idealized recording interaction.

The stochastic--quantum correspondence offers the prospect of an ontic description of quantum phenomena grounded in the principles of stochastic processes. The present work further develops and strengthens this proposal by demonstrating how phase-sensitive quantum predictions can be realized through ordinary probabilities over definite configuration trajectories, with the relevant predictive information encoded in context-dependent stochastic processes whose conditional probabilities may depend on configuration history. The quantum lift provides a compact representation of this predictive information without uniquely determining the underlying stochastic trajectory law.

\appendix
\renewcommand{\theequation}{\thesection.\arabic{equation}}
\section{Faithful configuration recording}
\label{app:intermediate-recording}

The rooted quantum probabilities of the idealized interaction in Eq.~\eqref{eq:configuration-recording} admit different stochastic realizers. We first give an example in which the device does not record the incoming system configuration, then show that a faithful realizer that does record it can always be chosen. The construction establishes compatibility with the rooted probabilities without selecting a unique trajectory law.

Use the notation $t_m^-,t_m^+,S_-,D_-,S_+,D_+$ introduced in Section~\ref{sec:faithful-recording}, and let $t_0<t_m^-$ be the original root. Write $S_0,D_0$ for the configurations at $t_0$. The device remains ready before recording, so $D_0=D_-=0$. Let $U$ describe the preceding system evolution, with initial lifted state $J(p_0)$.

\subsection{The rooted quantum predictions}

The isolated shadow matrix $M_{SD}=[C_{SD}]_{\odot}$ has the ready-device columns given in Eq.~\eqref{eq:recording-appendix-kernel}. When recording follows the earlier evolution $U$, the system--device kernel rooted at $t_0$ is prescribed by
\begin{equation}
	\Gamma_{SD}(t_m^+\leftarrow t_0)=[C_{SD}(U\otimes I_D)]_{\odot}.
	\label{eq:recording-appendix-composite-shadow}
\end{equation}
The corresponding rooted system kernel, evaluated immediately before the measurement interaction, is $\G{t_m^-}{t_0}=[U]_{\odot}$. The system--device rooted conditional probabilities are given by
\begin{equation}
	\begin{aligned}
		\text{At }t_m^-:\quad
		\Pr(S_-=r,D_-=e\mid S_0=i,D_0=0)
		&=[\G{t_m^-}{t_0}]_{ri}\delta_{e,0},\\
		\text{At }t_m^+:\quad
		\Pr(S_+=s,D_+=d\mid S_0=i,D_0=0)
		&=[\G{t_m^-}{t_0}]_{si}\delta_{d,d(s)},
	\end{aligned}
	\label{eq:recording-appendix-rooted-probabilities}
\end{equation}
where the second line follows by applying Eq.~\eqref{eq:recording-appendix-output-probabilities} with $\rho(t_m^-)=U\hat P^iU^\dagger$ for each sharp initial preparation $\hat P^i$.

Every realizer of these rooted kernels therefore satisfies
\[
\Pr(D_+=d(s))=\Pr(S_-=s)=\Pr(S_+=s),
\]
and $D_+=d(S_+)$ with probability one. These statements fix the relevant single-time distributions and the final system--device correlation, but do not determine the joint distribution of $S_-$ and $S_+$.

\subsection{An unfaithful completion with the same rooted kernels}

Take a two-state system with preceding evolution $U=H$, where
\[
H=\frac{1}{\sqrt2}
\begin{pmatrix}1&1\\1&-1\end{pmatrix},
\]
and a two-state device with $d(r)=r$. For either initial configuration $i=0,1$, Eq.~\eqref{eq:recording-appendix-rooted-probabilities} gives pre-recording system probabilities $1/2,1/2$ and post-recording system--device pairs $(0,0),(1,1)$ with probability $1/2$ each.

Now choose a trajectory law that samples $S_-$ uniformly, flips it, and records the final configuration. Write $K_{SD}$ for the flip-and-record matrix, whose ready-device columns are 
\begin{equation}
	[K_{SD}]_{(s,d),(r,0)}
	:=\delta_{s,1-r}\delta_{d,s},
	\qquad r,s,d\in\{0,1\}.
	\label{eq:recording-appendix-bad-kernel}
\end{equation}
The realizer's actual two-time kernel across the recording interval satisfies
\[
\begin{aligned}
	\bigl[\Gamma_{\mathrm{bad},SD}(t_m^+\leftarrow t_m^-)\bigr]_{(s,d),(r,0)}
	&=\Pr_{\mathrm{bad}}(S_+=s,D_+=d\mid S_-=r,D_-=0)\\
	&=[K_{SD}]_{(s,d),(r,0)}.
\end{aligned}
\]
With $D_0=D_-=0$, the corresponding joint law factors as
\begin{equation}
	\begin{aligned}
		&\Pr_{\mathrm{bad}}(S_0=i,S_-=r,S_+=s,D_+=d)\\
		&\qquad=p_0(i)[\G{t_m^-}{t_0}]_{ri}
		[K_{SD}]_{(s,d),(r,0)}\\
		&\qquad=\frac{p_0(i)}{2}
		\delta_{s,1-r}\delta_{d,s}.
	\end{aligned}
	\label{eq:recording-appendix-bad-law}
\end{equation}
Its rooted marginals are
\begin{equation}
	\begin{aligned}
		\text{At }t_m^-:\quad
		\Pr_{\mathrm{bad}}(S_-=r,D_-=e\mid S_0=i,D_0=0)
		&=\sum_{s,d}\frac12
		\delta_{s,1-r}\delta_{d,s}\delta_{e,0}
		=\frac12\delta_{e,0},\\
		\text{At }t_m^+:\quad
		\Pr_{\mathrm{bad}}(S_+=s,D_+=d\mid S_0=i,D_0=0)
		&=\sum_r\frac12
		\delta_{s,1-r}\delta_{d,s}
		=\frac12\delta_{d,s}.
	\end{aligned}
	\label{eq:recording-appendix-bad-marginals}
\end{equation}
These agree with both rooted quantum kernels for both initial configurations, and hence for every initial distribution. Nevertheless, $S_+=1-S_-$: the device always agrees with the final configuration and always disagrees with the preceding one.

The two realizers give the same rooted probabilities because the uniform pre-recording distribution is unchanged by the flip. Consequently, the matrices $K_{SD}$ and $M_{SD}$ act identically on the prescribed incoming distributions:
\begin{equation}
	\begin{aligned}
		\sum_r[K_{SD}]_{(s,d),(r,0)}
		[\G{t_m^-}{t_0}]_{ri}
		&=\frac12\delta_{d,s}\\
		&=\sum_r[M_{SD}]_{(s,d),(r,0)}
		[\G{t_m^-}{t_0}]_{ri}.
	\end{aligned}
	\label{eq:recording-appendix-bad-factorization}
\end{equation}
Thus even a factorization of the rooted kernels through $M_{SD}$ does not establish that the realizer's actual recording kernel has the ready-device columns of $M_{SD}$. 

\subsection{Existence of a faithful realizer}

Faithful recording, as defined in Eq.~\eqref{eq:faithful-recording}, amounts to choosing a realizer whose actual two-time kernel across the recording interval satisfies 
\begin{equation}
	\begin{aligned}
		\bigl[\Gamma_{SD}(t_m^+\leftarrow t_m^-)\bigr]_{(s,d),(r,0)}
		&=\Pr(S_+=s,D_+=d\mid S_-=r,D_-=0)\\
		&=[M_{SD}]_{(s,d),(r,0)}
		=\delta_{s,r}\delta_{d,d(r)},
	\end{aligned}
	\label{eq:recording-appendix-actual-transition}
\end{equation}
on conditioning events of positive probability. Here $r$ labels the immediately preceding configuration, rather than the original configuration $i$. We now construct such a realizer. Choose any stochastic realizer of the evolution preceding the configuration recording time, with the device ready throughout. For any finite collection of times $t_0<\cdots<t_m=t_m^-$, let $p_{\mathrm{pre}}(s_0,\ldots,s_m)$ be its joint distribution of system configurations. Extend it by
\begin{equation}
	\begin{aligned}
		p_{\mathrm{rec}}(s_0,\ldots,s_m,s,d)
		&=p_{\mathrm{pre}}(s_0,\ldots,s_m)\\
		&\qquad\times
		\delta_{s,s_m}\delta_{d,d(s_m)},
	\end{aligned}
	\label{eq:recording-appendix-path-extension}
\end{equation}
where $s$ and $d$ are the system and device configurations at $t_m^+$. Equivalently, set $S_+:=S_-$ and $D_+:=d(S_-)$ on each trajectory. This deterministic extension defines consistent joint laws and leaves the preceding path law unchanged. Its added factor is independent of $p_0$, so the conditional path laws given the initial configuration remain fixed as $p_0$ varies. In particular, the resulting three-time law is
\begin{equation}
	\begin{aligned}
		&\Pr(S_0=i,S_-=r,S_+=s,D_+=d)\\
		&\qquad=p_0(i)[\G{t_m^-}{t_0}]_{ri}
		\delta_{s,r}\delta_{d,d(r)},
	\end{aligned}
	\label{eq:recording-appendix-faithful-law}
\end{equation}
with $D_0=D_-=0$. Summing over $(s,d)$ or over $r$ reproduces, respectively, the pre-recording and post-recording rooted kernels in Eq.~\eqref{eq:recording-appendix-rooted-probabilities} for every initial distribution. Compatibility does not require a diagonal pre-recording lifted state: Eq.~\eqref{eq:recording-appendix-output-probabilities} holds for every $\rho(t_m^-)$, in particular for each $U\hat P^iU^\dagger$. The device outcome probabilities are
\begin{equation}
	\Pr(D_+=d(r))
	=\Pr(S_-=r)
	=\langle r|\rho(t_m^-)|r\rangle.
	\label{eq:recording-appendix-device-statistics}
\end{equation}

Finally, we note that faithfulness also fixes the recording result after conditioning on an earlier history. On conditioning events of positive probability,
\begin{equation}
	\begin{aligned}
		&\Pr(S_+=s,D_+=d\mid S_-=r,D_-=0,S_0=i,D_0=0)\\
		&\qquad=\Pr(S_+=s,D_+=d\mid S_-=r,D_-=0)
		=\delta_{s,r}\delta_{d,d(r)}.
	\end{aligned}
	\label{eq:recording-appendix-three-time-condition}
\end{equation}
The same conclusion holds for any earlier finite configuration history.\footnote{This follows from the deterministic two-time condition: the probability of correct recording is a weighted average of history-dependent probabilities, each at most one; if the average is one, every history with positive weight must also give correct recording with probability one. Earlier history may affect which configuration arrives, but cannot change the record once that configuration is given.}

\end{document}